\documentclass[conference,letter,romanappendices]{ieeeconf}
\let\proof\relax   

\usepackage{mathpple}
\usepackage{times}
\usepackage{amsthm,xpatch}
\usepackage{amsmath,amsfonts}
\usepackage{cite}
\usepackage{amssymb}
\usepackage{dsfont}
\usepackage{graphicx, subfigure}
\usepackage{color}
\usepackage{breqn}
\usepackage{mathtools}
\usepackage{bbm}
\usepackage{latexsym}
\usepackage[ruled, linesnumbered]{algorithm2e}
\usepackage{accents}
\usepackage{tikz}
\usepackage[mathscr]{euscript}
\usepackage{bm}
\usepackage[frak=lucida]{mathalpha} 
\usepackage{comment}

\usepackage[inline]{trackchanges}

\addeditor{}

\newtheorem{theorem}{Theorem}
\newtheorem{remark}{Remark}

  \renewcommand{\baselinestretch}{0.98}

\IEEEoverridecommandlockouts

\begin{document}

\newcommand{\SB}[3]{
\sum_{#2 \in #1}\biggl|\overline{X}_{#2}\biggr| #3
\biggl|\bigcap_{#2 \notin #1}\overline{X}_{#2}\biggr|
}

\newcommand{\Mod}[1]{\ (\textup{mod}\ #1)}

\newcommand{\overbar}[1]{\mkern 0mu\overline{\mkern-0mu#1\mkern-8.5mu}\mkern 6mu}

\makeatletter
\newcommand*\nss[3]{%
  \begingroup
  \setbox0\hbox{$\m@th\scriptstyle\cramped{#2}$}%
  \setbox2\hbox{$\m@th\scriptstyle#3$}%
  \dimen@=\fontdimen8\textfont3
  \multiply\dimen@ by 4             
  \advance \dimen@ by \ht0
  \advance \dimen@ by -\fontdimen17\textfont2
  \@tempdima=\fontdimen5\textfont2  
  \multiply\@tempdima by 4
  \divide  \@tempdima by 5          
  \ifdim\dimen@<\@tempdima
    \ht0=0pt                        
    \@tempdima=\fontdimen5\textfont2
    \divide\@tempdima by 4          
    \advance \dimen@ by -\@tempdima 
    \ifdim\dimen@>0pt
      \@tempdima=\dp2
      \advance\@tempdima by \dimen@
      \dp2=\@tempdima
    \fi
  \fi
  #1_{\box0}^{\box2}%
  \endgroup
  }
\makeatother

\makeatletter
\renewenvironment{proof}[1][\proofname]{\par
  \pushQED{\qed}%
  \normalfont \topsep6\p@\@plus6\p@\relax
  \trivlist
  \item[\hskip\labelsep
        \itshape
    #1\@addpunct{:}]\ignorespaces
}{%
  \popQED\endtrivlist\@endpefalse
}
\makeatother

\makeatletter
\newsavebox\myboxA
\newsavebox\myboxB
\newlength\mylenA

\newcommand*\xoverline[2][0.75]{%
    \sbox{\myboxA}{$\m@th#2$}%
    \setbox\myboxB\null
    \ht\myboxB=\ht\myboxA%
    \dp\myboxB=\dp\myboxA%
    \wd\myboxB=#1\wd\myboxA
    \sbox\myboxB{$\m@th\overline{\copy\myboxB}$}
    \setlength\mylenA{\the\wd\myboxA}
    \addtolength\mylenA{-\the\wd\myboxB}%
    \ifdim\wd\myboxB<\wd\myboxA%
       \rlap{\hskip 0.5\mylenA\usebox\myboxB}{\usebox\myboxA}%
    \else
        \hskip -0.5\mylenA\rlap{\usebox\myboxA}{\hskip 0.5\mylenA\usebox\myboxB}%
    \fi}
\makeatother

\xpatchcmd{\proof}{\hskip\labelsep}{\hskip3.75\labelsep}{}{}

\pagestyle{empty}

\title{\fontsize{21}{28}\selectfont The Linear Capacity of Single-Server Individually-Private Information Retrieval with Side Information}

\author{Anoosheh Heidarzadeh and Alex Sprintson\thanks{The authors are with the Department of Electrical and Computer Engineering, Texas A\&M University, College Station, TX 77843 USA (E-mail: \{anoosheh, spalex\}@tamu.edu).}
}

%

%


\maketitle 

\thispagestyle{empty}

\begin{abstract}
This paper considers the problem of single-server Individually-Private Information Retrieval with side information (IPIR). 
In this problem, there is a remote server that stores a dataset of $K$ messages, and there is a user that initially knows $M$ of these messages, and wants to retrieve $D$ other messages belonging to the dataset. 
The goal of the user is to retrieve the $D$ desired messages by downloading the minimum amount of information from the server while revealing no information about whether an individual message is one of the $D$ desired messages. 
In this work, we focus on linear IPIR schemes, i.e., the IPIR schemes in which the user downloads only linear combinations of the original messages from the server. 
We prove a converse bound on the download rate of any linear IPIR scheme for all $K,D,M$, and 
show the achievability of this bound for all $K,D,M$ satisfying a certain divisibility condition.
Our results characterize the linear capacity of IPIR, which is defined as the maximum achievable download rate over all linear IPIR schemes, for a wide range of values of $K,D,M$. 
\end{abstract}

\section{Introduction}
In this work, we consider the problem of \emph{single-server Individually-Private Information Retrieval with side information}, which we refer to as \emph{IPIR} for short.
In this problem, there is a set of $K$ messages stored on a remote server, and there is a user that has $M$ (out of $K$) messages as side information, and wants to retrieve $D$ other messages. 
The objective is to design a retrieval scheme in which the user downloads the minimum possible amount of information from the server while revealing no information about the identity of every individual message required by the user. 

The IPIR problem, which was originally introduced in~\cite{HKRS2019} and later studied in~\cite{HS2021}, is related to several work in the Private Information Retrieval (PIR) literature. 
In particular, the IPIR problem is a variant of the problem of multi-message PIR with side information (MPIR-SI) which is a generalization of the multi-message PIR problem~\cite{BU17,BU2018}.  
In the MPIR-SI problem, a user wishes to privately retrieve multiple messages, with the help of a prior side information, from a single (or multiple) remote server(s) storing (identical copies or coded versions of) a set of messages.

The MPIR-SI problem has been studied under three different information-theoretic privacy guarantees: 
\emph{full privacy}, \emph{joint privacy}, and \emph{individual privacy}.
In the case of full privacy, 
both the identities of the messages required by the user and the identities of the user's side information messages must be kept private from the server(s).
In contrast, when joint or individual privacy is required, only the identities of the messages required by the user must be protected, and it is not required to protect the identities of the user's side information messages. 
In the case of joint privacy, the server(s) must not learn which subset of messages was required by the user, whereas in the case of individual privacy, the server(s) must not learn whether an individual message was one of the messages required by the user. 

The joint privacy guarantee finds application in scenarios in which the correlation between the identities of the required messages must be protected (see, e.g.,~\cite{MMM2019,HES2022}), whereas the individual privacy guarantee is of practical importance in scenarios in which there is no need to protect the correlation between the identities of the required messages (see, e.g.,~\cite{KKHS32019,HES2021IndividualJournal}).
Note that the joint and individual privacy requirements are equivalent in the classical PIR problem and the problem of PIR with side information where the user wants to privately retrieve only one message~\cite{SJ2017,T2017,TER2017,TGKHHER2017,SJ2018No2,WBU2018,WBU2018No2,TSC2019,KGHERS2020,HKS2018,HKS2019,HKS2019Journal,KKHS12019,KKHS22019,CWJ2020,LG2020CISS,SS2021,KH2021}.


Several variants of the MPIR-SI problem were previously studied in the literature. 
The single-server setting of MPIR-SI with full privacy was studied in~\cite{HKGRS2018}, and the multi-server setting of this problem was studied in~\cite{SSM2018}.
In addition, the single-server setting of MPIR-SI with joint privacy, which we refer to as \emph{JPIR} for short, was studied in~\cite{HKGRS2018,LG2018}, and the single-server setting of MPIR-SI with individual privacy, which is the IPIR problem, was considered in~\cite{HKRS2019,HS2021}. 

In~\cite{HKRS2019}, we proposed an IPIR scheme for all ${K,D,M}$, and showed that this scheme achieves a download rate higher than that of the JPIR schemes of~\cite{HKGRS2018} and \cite{LG2018}.
The optimality of the scheme of~\cite{HKRS2019} was also shown for ${D=2,M=1}$, but it was left open in general. 
Recently, in~\cite{HS2021}, we showed that the scheme of~\cite{HKRS2019} is not always optimal, and proposed an optimal IPIR scheme for ${D=2,M=2}$ that achieves a download rate higher than that of the scheme of~\cite{HKRS2019}.
Notwithstanding, the fundamental limits of the IPIR problem have remained unknown for all other values of $D,M$.

In this work, we focus on linear IPIR schemes, i.e., the user downloads only linear combinations of the original messages from the server. 
We prove a converse bound on the download rate of any linear IPIR scheme for all $K,D,M$, and show the achievability of this bound for all $K,D,M$ satisfying a certain divisibility condition. 
Our results characterize the optimal download rate of linear IPIR and show the sub-optimality of the IPIR scheme of~\cite{HKRS2019}, for a wide range of values of $K,D,M$. 
Our converse proof technique relies on a mix of combinatorial, algebraic, and information-theoretic arguments that are tailored to the single-server setting, linear retrieval schemes, and the individual privacy guarantee. 
In addition, our achievability scheme is based on randomized partitioning and maximum distance separable (MDS) codes. 


\section{Problem Setup}\label{sec:SN}
Throughout, we denote random variables and their realizations by bold-face symbols and regular symbols, respectively.
For any integer $i\geq 1$, we denote ${\{1,\dots,i\}}$ by ${[i]}$. 
Also, we denote the binomial coefficient $\binom{n}{k}$ by ${C_{n,k}}$.
 
Let $\mathbbmss{F}_q$ be a finite field of order $q$, and let $\mathbbmss{F}_{q}^{n}$ be the $n$-dimensional vector space over $\mathbbmss{F}_q$. 
Let $K,D,M$ be arbitrary integers such that $D\geq 2$, $M\geq 1$, and ${K\geq D+M}$. 

Consider a server that stores $K$ messages ${\mathrm{X}_1,\dots,\mathrm{X}_K}$, where $\mathrm{X}_i=[X_{i,1},\dots,X_{i,n}]\in \mathbbmss{F}_q^{n}$ for $i\in [K]$. 
Note that $X_{i,j}\in \mathbbmss{F}_q$ for all $i\in [K]$ and all $j\in [n]$.
We refer to $X_{i,1},\dots,X_{i,n}$ as the \emph{symbols} of the message $\mathrm{X}_i$. 
For simplifying the notation, we denote by $\mathrm{X}_{\mathcal{I}}$ the set of messages $\{\mathrm{X}_i: i\in \mathcal{I}\}$ for every ${\mathcal{I}\subseteq [K]}$. 

Let $\mathbbmss{W}$ be the set of all $D$-subsets of $[K]$, and let $\mathbbmss{S}$ be the set of all $M$-subsets of $[K]$. 
Consider a user who initially knows the $M$ messages $\mathrm{X}_{\mathrm{S}}$ for a given $\mathrm{S}\in \mathbbmss{S}$, and wishes to retrieve the $D$ messages $\mathrm{X}_{\mathrm{W}}$ for a given $\mathrm{W}\in \mathbbmss{W}$. 
To avoid degenerate cases, we assume that  $\mathrm{W}\cap \mathrm{S} = \emptyset$. 
We refer to $\mathrm{X}_{\mathrm{W}}$ as the \emph{demand}, $\mathrm{X}_{\mathrm{S}}$ as the \emph{side information}, $\mathrm{W}$ as the \emph{index set of the demand}, $\mathrm{S}$ as the \emph{index set of the side information}, $D$ as the \emph{size of the demand}, and $M$ as the \emph{size of the side information}.

In this work, we make the following assumptions:  
\begin{enumerate}
\item $\mathbf{X}_1,\dots,\mathbf{X}_K$ are independent and uniformly distributed over $\mathbbmss{F}_{q}^{n}$. 
Thus, ${H(\mathbf{X}_{\mathcal{I}})= |\mathcal{I}| B}$ for all ${\mathcal{I}\subseteq [K]}$, where $B:= n\log_2 q$ is the entropy of a message (in bits). 
\item $(\mathbf{W},\mathbf{S})$ and $\mathbf{X}_{1},\dots,\mathbf{X}_K$ are independent. 
\item The distribution of $\mathbf{S}$ is uniform over all $\mathrm{S}\in \mathbbm{S}$, and the conditional distribution of $\mathbf{W}$ given $\mathbf{S}=\mathrm{S}$ is uniform over all $\mathrm{W}\in \mathbbm{W}$ such that $\mathrm{W}\cap \mathrm{S} = \emptyset$. 
Thus, ${\mathbb{P}(\mathbf{W} = \mathrm{W},\mathbf{S}=\mathrm{S})}=1/C_{K,M}\times 1/C_{K-M,D}$ for all $(\mathrm{W},\mathrm{S})\in \mathbbmss{W}\times \mathbbmss{S}$ such that $\mathrm{W}\cap\mathrm{S}=\emptyset$. 
Moreover, $\mathbb{P}(i\in \mathbf{W})={\sum_{\mathrm{W}\in \mathbbmss{W}: i\in \mathrm{W}}\mathbb{P}(\mathbf{W} = \mathrm{W})} = \sum_{\mathrm{W}\in \mathbbmss{W}: i\in \mathrm{W}}\sum_{\mathrm{S}\in \mathbbmss{S}:\mathrm{W}\cap\mathrm{S}=\emptyset}\mathbb{P}(\mathbf{W} = \mathrm{W},\mathbf{S}=\mathrm{S})= C_{K-1,D-1}\times C_{K-D,M} \times 1/C_{K,M}\times 1/C_{K-M,D} = D/K$ for all $i\in [K]$. 
\item The demand's size $D$, the side information's size $M$, and the distribution of $(\mathbf{W},\mathbf{S})$ are initially known by the server, whereas the realization $(\mathrm{W},\mathrm{S})$ is initially unknown to the server.
\end{enumerate}

Given $(\mathrm{W},\mathrm{S})$, the user generates a query $\mathrm{Q}^{[\mathrm{W},\mathrm{S}]}$, and sends it to the server. 
The query $\mathrm{Q}^{[\mathrm{W},\mathrm{S}]}$ is a deterministic or stochastic function of $(\mathrm{W},\mathrm{S})$, independent of $\mathrm{X}_1,\dots,\mathrm{X}_K$. 
Given the query, every message index must be equally likely to belong to the demand's index set. 
That is, for every $i\in [K]$, we must have
${\mathbb{P}(i\in \mathbf{W}|\mathbf{Q}^{[\mathbf{W},\mathbf{S}]}=\mathrm{Q}^{[\mathrm{W},\mathrm{S}]})=\mathbb{P}(i \in \mathbf{W}) = D/K}$. 
We refer to this condition as the \emph{individual privacy condition}.

Upon receiving $\mathrm{Q}^{[\mathrm{W},\mathrm{S}]}$, the server generates an answer $\mathrm{A}^{[\mathrm{W},\mathrm{S}]}$, and sends it back to the user. 
The answer $\mathrm{A}^{[\mathrm{W},\mathrm{S}]}$ is a deterministic function of $\mathrm{Q}^{[\mathrm{W},\mathrm{S}]}$ and $\mathrm{X}_1,\dots,\mathrm{X}_K$. 
That is, ${H(\mathbf{A}^{[\mathbf{W},\mathbf{S}]}|\mathbf{Q}^{[\mathbf{W},\mathbf{S}]},\mathbf{X}_1,\dots,\mathbf{X}_K)=0}$. 
The user must be able to recover the demand $\mathrm{X}_{\mathrm{W}}$ given the answer $\mathrm{A}^{[\mathrm{W},\mathrm{S}]}$, the query $\mathrm{Q}^{[\mathrm{W},\mathrm{S}]}$, the side information $\mathrm{X}_{\mathrm{S}}$, and the realization $(\mathrm{W}, \mathrm{S})$. 
That is,
$H(\mathbf{X}_{\mathrm{W}}| \mathbf{A}^{[\mathbf{W},\mathbf{S}]},\mathbf{Q}^{[\mathbf{W},\mathbf{S}]},\mathbf{X}_{\mathrm{S}})=0$. 
We refer to this condition as the \emph{recoverability condition}. 

The problem is to design a protocol for generating a query $\mathrm{Q}^{[\mathrm{W},\mathrm{S}]}$ and the corresponding answer $\mathrm{A}^{[\mathrm{W},\mathrm{S}]}$ for any given $(\mathrm{W},\mathrm{S})$
such that both the individual privacy and recoverability conditions are satisfied.
We refer to this problem as \emph{single-server Individually-Private Information Retrieval with side information}, or \emph{IPIR} for short. 

In this work, we focus on \emph{(scalar-) linear} protocols, i.e., any protocol in which the server's answer $\mathrm{A}^{[\mathrm{W},\mathrm{S}]}$ consists only of linear combinations of the messages $\mathrm{X}_1,\dots,\mathrm{X}_K$ (with combination coefficients from $\mathbbmss{F}_q$). 
Note that for any linear protocol, the user's query $\mathrm{Q}^{[\mathrm{W},\mathrm{S}]}$ can be fully specified by the corresponding combination coefficient vectors. 
Without loss of generality, we assume that these combination coefficient vectors are linearly independent (over $\mathbbmss{F}_q$).

We define the \emph{rate} of a linear IPIR protocol as the ratio of the amount of information required by the user to the amount of information downloaded from the server, i.e., $H(\mathbf{X}_{\mathbf{W}})/H(\mathbf{A}^{[\mathbf{W},\mathbf{S}]})$, and define the \emph{linear capacity} of IPIR as the supremum of rates over all linear IPIR protocols.
By the assumptions (1) and (2), it can be shown that (i) $H(\mathbf{X}_{\mathbf{W}}) = DB$, and (ii) $H(\mathbf{A}^{[\mathbf{W},\mathbf{S}]}) = H(\mathbf{L}^{[\mathbf{W},\mathbf{S}]})+\mathbb{E}[\mathbf{L}^{[\mathbf{W},\mathbf{S}]}] B$, where $\mathbf{L}^{[\mathbf{W},\mathbf{S}]}$ represents the number of linear combinations that constitute the answer $\mathbf{A}^{[\mathbf{W},\mathbf{S}]}$.
In particular, if $\mathbf{L}^{[\mathbf{W},\mathbf{S}]}$ is a constant random variable taking only the value $L$ for some integer $L\geq 1$ (i.e., all realizations of $\mathbf{A}^{[\mathbf{W},\mathbf{S}]}$ consist of $L$ linear combinations), then ${H(\mathbf{A}^{[\mathbf{W},\mathbf{S}]}) = LB}$, and the rate of the protocol is $D/L$.
It should be noted that (i) and (ii) do not rely on the assumptions (3) and (4).

In this work, our goal is to characterize the linear capacity of IPIR in terms of the parameters $K,D,M$. 

\section{Main Results}\label{sec:MR}
This section summarizes our main results. Theorem~\ref{thm:LIPIRConv} provides an upper bound on the linear capacity of IPIR for all $K,D,M$, and Theorem~\ref{thm:LIPIRAch} provides a matching lower bound on the linear capacity of IPIR for all $K,D,M$ such that $M$ is an integer multiple of $D$.
The proof of converse and achievability are presented in Sections~\ref{sec:Conv} and~\ref{sec:Ach}, respectively.

\begin{theorem}\label{thm:LIPIRConv}
For IPIR with $K$ messages, demand's size $D$, and side information's size $M$, the linear capacity is upper bounded by $(D+M)/K$.
\end{theorem}

We present a novel technique to prove the converse when the number of symbols of a message ($n$) grows unbounded. 
It should be noted that the converse bound holds for any $n\geq 1$. 
This is because the linear capacity for any finite $n$ cannot exceed the linear capacity as $n$ tends to infinity. (Any linear protocol for $n=n_0$ can serve as a linear protocol---achieving the same rate---for $n = k n_0$ for any arbitrary integer $k\geq 1$.)
Our proof technique is based on a mix of combinatorial, algebraic, and information-theoretic arguments which rely on the individual-privacy and recoverability conditions. 
The main idea of the proof is to show that given the answer and the query of any linear IPIR protocol, there exists a collection of coded symbols (each coded symbol being a linear combination of the symbols of a message with combination coefficients from $\mathbbmss{F}_q$) of size at most ${(M/(D+M))Kn}$ from which all $Kn$ message symbols can be recovered. 
To prove this, we use a standard random linear network coding argument and constructively identify the number of required coded symbols for each message. 
This result implies that the amount of information downloaded from the server in any linear IPIR protocol is lower bounded by ${KB-(M/(D+M))KB} = {(D/(D+M))KB}$ bits, where ${B = n\log_2 q}$ is the amount of information in a message (in bits). 
Since the amount of information required by the user is $DB$ bits, then the rate of any linear IPIR protocol is upper bounded by ${DB/((D/(D+M))KB)} = (D+M)/K$.

\begin{theorem}\label{thm:LIPIRAch}
For IPIR with $K$ messages, demand's size $D$, and side information's size $M$, the linear capacity is lower bounded by $(D+M)/K$ when $(D+M)\mid K$, or more generally, when $((D+M)/\gcd(D,M))\mid K$. 
\end{theorem}

To prove the achievability result, we propose a capacity-achieving linear IPIR protocol, which we call \emph{Group-and-Code}, for all ${K,D,M}$ such that ${D/R+M/R}$ is an integer divisor of $K$, where ${R:= \gcd(D,M)}$. 
This protocol is applicable for any ${n\geq 1}$ and any ${q\geq 2}$ such that there exists a $[D/R+M/R,D/R]$ MDS code over $\mathbbmss{F}_q$ (e.g., any ${q\geq 2}$ or any ${q\geq D/R+M/R}$ when ${D/R=1}$ or ${D/R>1}$, respectively.)
The key idea of the proposed protocol is
to carefully divide the messages into ${K/(D/R+M/R)}$ disjoint groups of size ${D/R+M/R}$, and query $D/R$ MDS-coded combinations of the messages in each group.   


\begin{remark}\label{rem:LIPIRCap}
\emph{The results of Theorems~\ref{thm:LIPIRConv} and~\ref{thm:LIPIRAch} show that the linear capacity of IPIR is given by $(D+M)/K$ for all $K,D,M$ such that $D/R+M/R$ divides $K$. 
The highest rate previously shown to be achievable for IPIR~\cite{HKRS2019} was ${D/(K-M\lfloor K/{(D+M)}\rfloor)}$ for all $K,D,M$ such that ${{(K-D)}/{(D+M)}\leq \lfloor {K}/{(D+M)}\rfloor}$, and ${1/\lceil {K}/{(D+M)} \rceil}$ for all $K,D,M$ such that ${{(K-D)}/{(D+M)}> \lfloor {K}/{(D+M)}\rfloor}$. 
This achievable rate reduces to $(D+M)/K$ when $(D+M)\mid K$, yet it is strictly lower than $(D+M)/K$ when $(D+M)\nmid K$.
In contrast, the proposed scheme in this work achieves the rate $(D+M)/K$ for a wider range of values of $K,D,M$, particularly when $(D+M)\nmid K$ but $(D/R+M/R) \mid K$. 
Not only does this show the sub-optimality of the IPIR scheme of~\cite{HKRS2019} in general, but it also shows the significance of the results in this work. 
We conjecture that the linear capacity of IPIR is given by $D/\lceil DK/(D+M)\rceil$ for all $K,D,M$, which reduces to $(D+M)/K$ when $(D/R+M/R)\mid K$.}
\end{remark}

\begin{remark}\label{rem:IPIRCap}
\emph{The (general) capacity of IPIR, which is defined as the maximum achievable rate over all linear and non-linear IPIR protocols, was previously characterized for the two cases of ${D=2,M=1}$ and ${D=2,M=2}$ in~\cite{HKRS2019} and~\cite{HS2021}, respectively. 
In particular, the capacity was shown to be ${2/\lceil 2K/3 \rceil}$ and ${2/\lceil K/2\rceil}$ for ${D=2,M=1}$ and ${D=2,M=2}$, respectively. 
In addition, it was shown that in both of these cases the capacity can be achieved by a linear protocol.  
This shows that the linear capacity of IPIR and the general capacity of IPIR for these cases are the same.  
We conjecture that the general capacity of IPIR does not exceed the linear capacity of IPIR for any $K,D,M$, and it is given by ${D/\lceil DK/(D+M)\rceil}$ for all $K,D,M$.} 
\end{remark}

\begin{remark}\label{rem:IPIRAsymp}
\emph{The linear capacity of IPIR, which is given by ${(D+M)/K}$ for all $K,D,M$ such that $(D/R+M/R) \mid K$, may approach zero as $K$ tends to infinity, e.g., when $D$ and $M$ are constant with respect to $K$. 
However, the linear capacity of IPIR does not always approach zero as $K$ grows unbounded. 
For instance, when $D=\alpha K$ and $M=\beta K$ for arbitrary constants $0<\alpha,\beta<1$ such that ${\alpha+\beta\leq 1}$ (i.e., $D$ and $M$ grow linearly in $K$), as $K$ grows unbounded, the linear capacity of IPIR approaches the nonzero constant ${\alpha+\beta}$, which can be as large as $1$.}
\end{remark}

\section{Proof of Theorem~\ref{thm:LIPIRConv}}\label{sec:Conv}
For simplifying the notation, we denote $\mathbf{Q}^{[\mathbf{W},\mathbf{S}]}$ and $\mathbf{A}^{[\mathbf{W},\mathbf{S}]}$ by $\mathbf{Q}$ and $\mathbf{A}$, respectively. 
We need to show that ${H(\mathbf{A})\geq (D/(D+M))KB}$, where ${B = n\log_2 q}$ is the entropy of a message (in bits). 
Recall that each message $\mathrm{X}_i$ consists of $n$ independent and uniformly distributed symbols ${\{{X}_{i,j}\}_{j\in [n]}}$ over $\mathbbmss{F}_q$. 
Let $N:=q^n-1$, and let ${\mathrm{c}_1,\dots,\mathrm{c}_{N}}$ be the set of all nonzero vectors in $\mathbbmss{F}_{q}^n$. 
For each ${m\in [N]}$, let ${\mathrm{c}_m=[c_{m,1},\dots,c_{m,n}]}$. 
Note that ${c_{m,j}\in \mathbbmss{F}_q}$ for all ${m\in [N]}$ and for all ${j\in [n]}$. 
We refer to ${Y_{i,m}:=\sum_{j\in [n]} c_{m,j}X_{i,j}}$ as a \emph{coded symbol} of the message $\mathrm{X}_i$. 
To prove that ${H(\mathbf{A})\geq (D/(D+M))KB}$, 
it suffices to show that for any linear IPIR protocol, given the query and the answer, there exist ${R\leq (M/(D+M))Kn}$ coded symbols ${\{Y_{i,m}\}_{\mathcal{T}}:=\{Y_{i,m}: (i,m)\in \mathcal{T}\}}$ 
for some $R$-subset $\mathcal{T}$ of ${[K]\times [N]}$ (depending on the realization of the query and the answer)
given which all $Kn$ message symbols ${\{X_{i,j}\}:=\{X_{i,j}: i\in [K], j\in [n]\}}$ can be recovered, 
i.e.,
${H(\{\mathbf{X}_{i,j}\}|\mathbf{A},\mathbf{Q},\{\mathbf{Y}_{i,m}\}_{\mathcal{T}}) = 0}$.
This is because 
\begin{align*}
H(\mathbf{A}) & \geq  H(\mathbf{A}|\mathbf{Q},\{\mathbf{Y}_{i,m}\}_{\mathcal{T}}) \nonumber \\ 
& \stackrel{\scriptsize\text{(a)}}{=} H(\mathbf{A}|\mathbf{Q},\{\mathbf{Y}_{i,m}\}_{\mathcal{T}}) + H(\{\mathbf{X}_{i,j}\}|\mathbf{A},\mathbf{Q},\{\mathbf{Y}_{i,m}\}_{\mathcal{T}})\\
& \stackrel{\scriptsize\text{(b)}}{=} H(\{\mathbf{X}_{i,j}\}|\mathbf{Q},\{\mathbf{Y}_{i,m}\}_{\mathcal{T}}) + H(\mathbf{A}|\mathbf{Q},\{\mathbf{X}_{i,j}\}) \\
& \stackrel{\scriptsize\text{(c)}}{=} H(\{\mathbf{X}_{i,j}\}|\mathbf{Q},\{\mathbf{Y}_{i,m}\}_{\mathcal{T}}) \\
& \stackrel{\scriptsize\text{(d)}}{=} H(\{\mathbf{X}_{i,j}\}|\{\mathbf{Y}_{i,m}\}_{\mathcal{T}}) \\
& \stackrel{\scriptsize\text{(e)}}{\geq} H(\{\mathbf{X}_{i,j}\}) - H(\{\mathbf{Y}_{i,m}\}_{\mathcal{T}}) \\
& \stackrel{\scriptsize\text{(f)}}{\geq} {(Kn-R)B/n} \\
& \stackrel{\scriptsize\text{(g)}}{\geq} (D/(D+M))KB, 
\end{align*} 
where (a) holds because ${H(\{\mathbf{X}_{i,j}\}|\mathbf{A},\mathbf{Q},\{\mathbf{Y}_{i,m}\}_{\mathcal{T}}) = 0}$ by assumption;
(b) follows from the chain rule of entropy;
(c) holds because $H(\mathbf{A}|\mathbf{Q},\{\mathbf{X}_{i,j}\}) = 0$;
(d) follows because $\mathbf{Q}$ and $\{\mathbf{X}_{i,j}\}$ are independent (by assumption);
(e) holds because $H(\{\mathbf{X}_{i,j}\}|\{\mathbf{Y}_{i,m}\}_{\mathcal{T}}) = H(\{\mathbf{X}_{i,j}\}) +H(\{\mathbf{Y}_{i,m}\}_{\mathcal{T}}|\{\mathbf{X}_{i,j}\}) - H(\{\mathbf{Y}_{i,m}\}_{\mathcal{T}})$, and ${H(\{\mathbf{Y}_{i,m}\}_{\mathcal{T}}|\{\mathbf{X}_{i,j}\}) = 0}$ because $Y_{i,m}$ is a linear combination of $\{X_{i,j}: j\in [n]\}$;
(f) holds because
${H(\{\mathbf{X}_{i,j}\}) = KB}$, 
noting that $\{\mathbf{X}_{i,j}\}$ are independent and uniformly distributed over $\mathbbmss{F}_q$, and
${H(\{\mathbf{Y}_{i,m}\}_{\mathcal{T}})\leq |\mathcal{T}| H(\mathbf{Y}_{i,m}) =  RB/n}$, 
noting that $\mathbf{Y}_{i,m}$ is a linear combination of ${\{\mathbf{X}_{i,j}: j\in [n]\}}$, and hence uniformly distributed over $\mathbbmss{F}_q$, 
i.e., ${H(\mathbf{Y}_{i,m}) = \log_2 q = B/n}$; 
and (g) follows because ${R\leq (M/(D+M))Kn}$ by assumption.

Consider an arbitrary linear IPIR protocol. 
Fix arbitrary ${\mathrm{W}\in \mathbbmss{W}}$ and ${\mathrm{S}\in \mathbbmss{S}}$ such that $\mathrm{W}\cap \mathrm{S} = \emptyset$. 
Let $\mathrm{Q}^{[\mathrm{W},\mathrm{S}]}$ and $\mathrm{A}^{[\mathrm{W},\mathrm{S}]}$ be a query and its corresponding answer generated by the protocol, respectively.
For ease of notation, we denote $\mathrm{Q}^{[\mathrm{W},\mathrm{S}]}$ and $\mathrm{A}^{[\mathrm{W},\mathrm{S}]}$ by $\mathrm{Q}$ and $\mathrm{A}$, respectively. 

For any ${(\mathrm{W}^{*},\mathrm{S}^{*})\in \mathbbmss{W}\times\mathbbmss{S}}$ such that ${\mathrm{W}^{*}\cap\mathrm{S}^{*} = \emptyset}$, 
we say that the tuple ${(\mathrm{W}^{*},\mathrm{S}^{*})}$ is \emph{feasible} (given $\mathrm{Q}$ and $\mathrm{A}$) if $\mathrm{X}_{\mathrm{W}^{*}}$ can be recovered given ${\mathrm{X}_{\mathrm{S}^{*}}}$.
Let ${(\mathrm{W}_1,\mathrm{S}_1),\dots,(\mathrm{W}_T,\mathrm{S}_T)}$ be the set of all feasible tuples given $\mathrm{Q}$ and $\mathrm{A}$.
By the linearity of the protocol, it follows that 
for any ${l\in [T]}$, any ${k\in \mathrm{W}_l}$, and any ${j\in [n]}$, the message symbol $X_{k,j}$ (i.e., the $j$th symbol of the message $\mathrm{X}_k$) can be recovered given the $M$ message symbols ${\{X_{i,j}\}_{i\in \mathrm{S}_l}}$. 
This further implies that for any ${k\in \mathrm{W}_l}$ and any ${m\in [N]}$, the coded symbol ${\sum_{j\in [n]} c_{m,j} X_{k,j}}$ can be recovered given the $M$ coded symbols ${\{\sum_{j\in [n]} c_{m,j} X_{i,j}\}_{i\in \mathrm{S}_l}}$. 

Recall that we need to show that there exist ${R\leq (M/(D+M))Kn}$ coded symbols given which all $Kn$ message symbols can be recovered.
As discussed in Section~\ref{sec:MR}, it suffices to prove this claim for sufficiently large $n$. 
In the following we present a proof by construction.

For each ${l\in [T]}$, 
let $p_l$ be the conditional probability that ${\mathbf{W} = \mathrm{W}_l}$ and ${\mathbf{S}=\mathrm{S}_l}$ given that ${\mathbf{Q}=\mathrm{Q}}$. 
That is,
${p_{l} := {\mathbb{P}(\mathbf{W}=\mathrm{W}_l,\mathbf{S}=\mathrm{S}_l|\mathbf{Q}=\mathrm{Q})}}$.
Note that ${\sum_{l\in [T]} p_l = 1}$.
For any ${i\in [K]}$, we define ${\mathcal{F}_{i}:= \{l\in [T]: i\in \mathrm{W}_l\}}$ and ${\mathcal{E}_{i}:= \{l\in [T]: i\in \mathrm{S}_l\}}$.
Let ${\alpha_i := \sum_{l\in \mathcal{F}_i} p_l}$ and ${\beta_i := \sum_{l\in \mathcal{E}_i} p_l}$ for all ${i\in [K]}$.
Note that ${\alpha_i
\stackrel{\tiny\text{(a)}}{=}\mathbb{P}(i\in \mathbf{W}|\mathbf{Q}=\mathrm{Q})\stackrel{\tiny\text{(b)}}{=}\mathbb{P}(i\in \mathbf{W}) = D/K}$, where (a) follows from the law of total probability; and (b) follows from the individual privacy condition.
To simplify the notation, we define $\alpha:=D/K$. 
Note also that ${\beta_i = \mathbb{P}(i\in \mathbf{S}|\mathbf{Q}=\mathrm{Q})}$ (by the law of total probability).
Moreover, ${\sum_{i\in [K]} \beta_i = M}$. 
This is because ${\sum_{i\in [K]} \beta_i = \sum_{i\in [K]} \sum_{l\in \mathcal{E}_i} p_l = \sum_{l\in [T]} M p_l = M}$. 

Let ${\mathcal{I}_0 = \emptyset}$, and let ${\mathcal{I}_1,\dots,\mathcal{I}_P}$ be a partition of $[K]$ such that 
for each ${k\in [P]}$, ${\beta_i = \beta_j}$ for all ${i,j\in \mathcal{I}_k}$, and 
for every ${k,h\in [P]}$, ${\beta_i\neq \beta_j}$ for all ${i\in \mathcal{I}_{k}}$ and for all ${j\in \mathcal{I}_{h}}$.  
For each ${k\in [P]}$, let $\gamma_k$ be such that ${\beta_i = \gamma_k}$ for all ${i\in \mathcal{I}_k}$. 
Without loss of generality, assume that ${\gamma_{1}> \gamma_2>\dots > \gamma_P}$.  
Let ${r_0:=0}$, and let ${r_k := n/(\alpha+\gamma_k)-\sum_{h=0}^{k-1} r_h}$ for all ${k\in [P]}$.

For sufficiently large $n$, we show that 
(i) there exist $\mathcal{N}_{k,l}\subset [N]$, ${|\mathcal{N}_{k,l}|=r_kp_l}$ for all ${k\in [P]}$ and for all ${l\in [T]}$, such that all $Kn$ message symbols ${\{X_{i,j}: i\in [K], j\in [n]\}}$ can be recovered given the coded symbols
${\{Y_{i,m}: i\in [K]\setminus \cup_{h=0}^{k-1}\mathcal{I}_h, m\in \mathcal{N}_{k,\mathcal{E}_i}\}_{k\in [P]}}$, and 
(ii) ${|\{Y_{i,m}: i\in [K]\setminus \cup_{h=0}^{k-1}\mathcal{I}_h, m\in \mathcal{N}_{k,\mathcal{E}_i}\}_{k\in [P]}|}\leq {(M/(D+M))Kn}$, where ${\mathcal{N}_{k,\mathcal{F}_i}:=\{\mathcal{N}_{k,l}\}_{l\in \mathcal{F}_i}}$ and ${\mathcal{N}_{k,\mathcal{E}_i}:=\{\mathcal{N}_{k,l}\}_{l\in \mathcal{E}_i}}$. 
From now on, whenever we use the notation $\mathcal{N}_{k,l}$ for any $k\in [P]$ and $l\in [T]$, it is assumed that $\mathcal{N}_{k,l}$ is a subset of $[N]$ of size $r_kp_l$, noting that ${r_kp_l}$ is an integer, for sufficiently large $n$.

We say that a collection  $\{\mathcal{N}_{1,l}\}_{l\in [T]}$ is \emph{good} if all message symbols ${\{X_{i,j}: i\in \mathcal{I}_1, j\in [n]\}}$ can be recovered given the coded symbols ${\{Y_{i,m}: i\in [K], m\in \mathcal{N}_{1,\mathcal{E}_i}\}}$. 
Also, we say that a collection $\{\mathcal{N}_{k,l}\}_{l\in [T]}$ for any $k\in [P]\setminus \{1\}$ is \emph{good} if given $k-1$ good collections $\{\mathcal{N}_{h,l}\}_{h\in [k-1],l\in [T]}$, all message symbols ${\{X_{i,j}:i\in \mathcal{I}_k, j\in [n]\}}$ can be recovered given the coded symbols ${\{Y_{i,m}: i\in [K]\setminus \cup_{g=0}^{h-1}\mathcal{I}_g, m\in \mathcal{N}_{h,\mathcal{E}_i}\}_{h\in [k]}}$. 

First, we consider the case of ${k=1}$, and prove the existence of a good collection ${\{\mathcal{N}_{1,l}\}_{l\in [T]}}$.
For any arbitrary collection ${\{\mathcal{N}_{1,l}\}_{l\in [T]}}$, it is easy to see that the coded symbols
${\{Y_{i,m}:i\in [K], m\in \mathcal{N}_{1,\mathcal{F}_i}\}}$ 
can be recovered given the coded symbols ${\{Y_{i,m}: i\in [K], m\in \mathcal{N}_{1,\mathcal{E}_i}\}}$.
By a standard random linear network coding argument~\cite{HMKKESL2006}, 
it can be shown that for sufficiently large $n$ (depending on $K$ and $q$), for randomly chosen ${\{\mathcal{N}_{1,l}\}_{l\in [T]}}$, 
the coded symbols ${\{Y_{i,m}:i\in [K], m\in \mathcal{N}_{1,\mathcal{F}_i}\cup\mathcal{N}_{1,\mathcal{E}_i}\}}$ are linearly independent combinations of the message symbols ${\{X_{i,j}\}}$ with a nonzero probability.
This implies that there exists a collection ${\{\mathcal{N}_{1,l}\}_{l\in [T]}}$ such that the coded symbols ${\{Y_{i,m}:i\in [K], m\in \mathcal{N}_{1,\mathcal{F}_i}\cup \mathcal{N}_{1,\mathcal{E}_i}\}}$ are linearly independent combinations of the message symbols ${\{X_{i,j}\}}$.
Fix such a collection ${\{\mathcal{N}_{1,l}\}_{l\in [T]}}$.
Note that ${|\{Y_{i,m}: m\in \mathcal{N}_{1,\mathcal{F}_i}\cup \mathcal{N}_{1,\mathcal{E}_i}\}|}={|\mathcal{N}_{1,\mathcal{F}_i}\cup \mathcal{N}_{1,\mathcal{E}_i}|}=n$ for all ${i\in \mathcal{I}_1}$. 
This is because for each ${i\in \mathcal{I}_1}$, we have
(i) $|\mathcal{N}_{1,\mathcal{F}_i}|=\sum_{l\in \mathcal{F}_i} |\mathcal{N}_{1,l}|= \sum_{l\in \mathcal{F}_i} r_1 p_l = r_1 \alpha$; 
(ii) $|\mathcal{N}_{1,\mathcal{E}_i}|=\sum_{l\in \mathcal{E}_i} |\mathcal{N}_{1,l}|= \sum_{l\in \mathcal{E}_i} r_1 p_l = r_1 \gamma_1$; 
(iii) $\mathcal{N}_{1,\mathcal{F}_i}$ and $\mathcal{N}_{1,\mathcal{E}_i}$ are disjoint 
since ${\{Y_{i,m}:m\in \mathcal{N}_{1,\mathcal{F}_i}\cup \mathcal{N}_{1,\mathcal{E}_i}\}}$ are linearly independent combinations of ${\{X_{i,j}: j\in [n]\}}$; and 
(iv) ${r_1\alpha+r_1\gamma_1=n}$ by the definition of $r_1$. 
Note that ${\{Y_{i,m}:m\in \mathcal{N}_{1,\mathcal{F}_i}\cup \mathcal{N}_{1,\mathcal{E}_i}\}}$ are $n$ linearly independent combinations of the message symbols ${\{\mathrm{X}_{i,j}:j\in [n]\}}$ for each ${i\in \mathcal{I}_1}$, and 
${\{Y_{i,m}: m\in \mathcal{N}_{1,\mathcal{F}_i}\}}$ can be recovered given ${\{Y_{i,m}: m\in \mathcal{N}_{1,\mathcal{E}_i}\}}$ for any ${i\in \mathcal{I}_1}$.
This implies that all message symbols ${\{X_{i,j}: i\in \mathcal{I}_1, j\in [n]\}}$ can be recovered given ${\{Y_{i,m}: i\in [K], m\in \mathcal{N}_{1,\mathcal{E}_i}\}}$. 
Thus, ${\{\mathcal{N}_{1,l}\}_{l\in [T]}}$ is a good collection. 
Note also that ${|\{Y_{i,m}: i\in [K], m\in \mathcal{N}_{1,\mathcal{E}_i}\}|}={\sum_{i\in [K]} |\mathcal{N}_{1,\mathcal{E}_i}|}={\sum_{i\in [K]} \sum_{l\in \mathcal{E}_i} r_1p_l}={\sum_{i\in [K]} r_1\beta_i}$.

Next, we show that for any ${k\in [P]\setminus \{1\}}$, given any $k-1$ good collections ${\{\mathcal{N}_{h,l}\}_{h\in [k-1],l\in [T]}}$, there exists a good collection ${\{\mathcal{N}_{k,l}\}_{l\in [T]}}$.
Consider the case of $k=2$. 
Fix a good collection ${\{\mathcal{N}_{1,l}\}_{l\in [T]}}$.
For any arbitrary collection ${\{\mathcal{N}_{2,l}\}_{l\in [T]}}$, the coded symbols ${\{Y_{i,m}:i\in [K], m\in \mathcal{N}_{2,\mathcal{F}_i}\}}$ can be recovered given the coded symbols ${\{Y_{i,m}: i\in [K], m\in \mathcal{N}_{2,\mathcal{E}_i}\}}$.
Recall that ${\{X_{i,j}: i\in \mathcal{I}_1, j\in [n]\}}$ can be recovered given ${\{Y_{i,m}: i\in [K], m\in \mathcal{N}_{1,\mathcal{E}_i}\}}$. 
This implies that the coded symbols
${\{Y_{i,m}: i\in \mathcal{I}_1, m\in \mathcal{N}_{2,\mathcal{E}_i}\}}$ can be recovered given the coded symbols ${\{Y_{i,m}: i\in [K], m\in \mathcal{N}_{1,\mathcal{E}_i}\}}$.
By combining these arguments, 
one can observe that the coded symbols ${\{Y_{i,m}: i\in [K], m\in \mathcal{N}_{1,\mathcal{F}_i}\}}$ and the coded symbols ${\{Y_{i,m}: i\in [K], m\in \mathcal{N}_{2,\mathcal{F}_i}\}}$ can be recovered given the coded symbols ${\{Y_{i,m}: i\in [K], m\in \mathcal{N}_{1,\mathcal{E}_i}\}}$ and the coded symbols ${\{Y_{i,m}: i\in [K]\setminus \mathcal{I}_1, m\in \mathcal{N}_{2,\mathcal{E}_i}\}}$. 
Similarly as before, it can be shown that given a good collection ${\{\mathcal{N}_{1,l}\}_{l\in [T]}}$, 
there exists a collection  ${\{\mathcal{N}_{2,l}\}_{l\in [T]}}$ such that the coded symbols ${\{Y_{i,m}: i\in [K], m\in \mathcal{N}_{1,\mathcal{F}_i}\}}$, ${\{Y_{i,m}: i\in [K], m\in  \mathcal{N}_{1,\mathcal{E}_i}\}}$, ${\{Y_{i,m}: i\in [K], m\in  \mathcal{N}_{2,\mathcal{F}_i}\}}$, and ${\{Y_{i,m}: i\in [K]\setminus \mathcal{I}_1, m\in  \mathcal{N}_{2,\mathcal{E}_i}\}}$ are linearly independent combinations of the message symbols $\{X_{i,j}\}$.
Fix such a collection ${\{\mathcal{N}_{2,l}\}_{l\in [T]}}$. 
By the same arguments as before, 
${|\{Y_{i,m}: m\in \mathcal{N}_{1,\mathcal{F}_i}\cup \mathcal{N}_{1,\mathcal{E}_i}\cup \mathcal{N}_{2,\mathcal{F}_i}\cup \mathcal{N}_{2,\mathcal{E}_i}\}|}={|\mathcal{N}_{1,\mathcal{F}_i}|+| \mathcal{N}_{1,\mathcal{E}_i}|+|\mathcal{N}_{2,\mathcal{F}_i}|+ |\mathcal{N}_{2,\mathcal{E}_i}|}={ (r_1+r_2)(\gamma_2+\alpha)}={n}$ for all ${i\in \mathcal{I}_2}$. 
It is easy to verify that for each ${i\in \mathcal{I}_2}$, the coded symbols ${\{Y_{i,m}:m\in \mathcal{N}_{1,\mathcal{F}_i}\cup \mathcal{N}_{1,\mathcal{E}_i}\cup \mathcal{N}_{2,\mathcal{F}_i}\cup \mathcal{N}_{2,\mathcal{E}_i}\}}$ are $n$ linearly independent combinations of the message symbols $\{\mathrm{X}_{i,j}: j\in [n]\}$. 
Moreover, as we showed earlier, the coded symbols ${\{Y_{i,m}: i\in [K], m\in \mathcal{N}_{1,\mathcal{F}_i}\}}$ and 
${\{Y_{i,m}: i\in [K], m\in \mathcal{N}_{2,\mathcal{F}_i}\}}$ can be recovered given the coded symbols ${\{Y_{i,m}: i\in [K], m\in \mathcal{N}_{1,\mathcal{E}_i}\}}$ and ${\{Y_{i,m}: i\in [K]\setminus \mathcal{I}_1, m\in \mathcal{N}_{2,\mathcal{E}_i}\}}$. 
Putting these arguments together, it follows that all message symbols ${\{X_{i,j}: i\in \mathcal{I}_2, j\in [n]\}}$ can be recovered given the coded symbols
${\{Y_{i,m}: i\in [K]\setminus \cup_{g=0}^{h-1} \mathcal{I}_g, m\in \mathcal{N}_{h,\mathcal{E}_i}\}_{h\in [2]}}$.
This implies that ${\{\mathcal{N}_{2,l}\}_{l\in [T]}}$ is a good collection. 
Note also that 
${|\{Y_{i,m}: i\in [K]\setminus \cup_{g=0}^{h-1} \mathcal{I}_g, m\in \mathcal{N}_{h,\mathcal{E}_i}\}_{h\in [2]}|}={\sum_{i\in [K]} |\mathcal{N}_{1,\mathcal{E}_i}|}+{\sum_{i\in [K]\setminus \mathcal{I}_1} |\mathcal{N}_{2,\mathcal{E}_i}|} = {\sum_{i\in [K]} r_1\beta_i}+{\sum_{i\in [K]\setminus \mathcal{I}_1} r_2\beta_i} = {\sum_{i\in \mathcal{I}_1}r_1\beta_i}+ {\sum_{i\in [K]\setminus\mathcal{I}_1}(r_1+r_2)\beta_i}$. 

Repeating the same arguments as above for the cases of $k=3,\dots,P$, it can be shown that for each ${k\in [P]\setminus \{1\}}$, given any $k-1$ good collections ${\{\mathcal{N}_{h,l}\}_{h\in [k-1],l\in [T]}}$, there exists a good collection ${\{\mathcal{N}_{k,l}\}_{l\in [T]}}$. 
That is, there exist ${\{\mathcal{N}_{k,l}\}_{l\in [T]}}$ such that all message symbols ${\{X_{i,j}: i\in \mathcal{I}_k, j\in [n]\}}$ can be recovered given the coded symbols ${\{Y_{i,m}: i\in [K]\setminus \cup_{g=0}^{h-1}\mathcal{I}_g, m\in \mathcal{N}_{h,\mathcal{E}_i}\}_{h\in [k]}}$.
This implies that for each $k\in [P]$, there exist ${\{\mathcal{N}_{h,l}\}_{h\in [k],l\in [T]}}$ such that all message symbols ${\{X_{i,j}: i\in \cup_{h=1}^{k}\mathcal{I}_h, j\in [n]\}}$ can be recovered given the coded symbols
${\{Y_{i,m}: i\in [K]\setminus \cup_{g=0}^{h-1} \mathcal{I}_g, m\in \mathcal{N}_{h,\mathcal{E}_i}\}_{h\in [k]}}$. 
Using the same arguments as before, it can be shown that for each $k\in [P]$, ${|\{Y_{i,m}: i\in [K]\setminus \cup_{g=0}^{h-1} \mathcal{I}_{g}, m\in \mathcal{N}_{h,\mathcal{E}_i}\}_{h\in [k]}|}={\sum_{h\in [k]} \sum_{i\in [K]\setminus (\mathcal{I}_0\cup \dots \cup \mathcal{I}_{h-1})} r_h \beta_i}$. 
Taking $k=P$, it then follows that 
all $Kn$ message symbols $\{X_{i,j}\}$ can be recovered given the $\sum_{h\in [P]} \sum_{i\in [K]\setminus (\mathcal{I}_0\cup \dots \cup \mathcal{I}_{h-1})} r_h \beta_i$ coded symbols
${\{Y_{i,m}: i\in [K]\setminus \cup_{g=0}^{h-1} \mathcal{I}_{g}, m\in \mathcal{N}_{h,\mathcal{E}_i}\}_{h\in [P]}}$.
It is also easy to verify that ${\sum_{h\in [P]} \sum_{i\in [K]\setminus (\mathcal{I}_0\cup \dots \cup \mathcal{I}_{h-1})} r_h \beta_i}={\sum_{h\in [P]} \sum_{i\in \mathcal{I}_h} (r_1+\dots+r_h)\beta_i} = {(\sum_{i\in [K]} (\beta_i/(\alpha+\beta_i)))n}$.
By combining these arguments, it follows that there exist ${(\sum_{i\in [K]} (\beta_i/(\alpha+\beta_i)))n}$ coded symbols given which all $Kn$ message symbols can be recovered.

To complete the proof, we need to show that ${(\sum_{i\in [K]} (\beta_i/(\alpha+\beta_i)))n\leq (M/(D+M))kn}$.
It is easy to show that ${\sum_{i\in [K]} (\beta_i/(\alpha+\beta_i))}$ is maximized when ${\beta_i = M/K}$ for all $i\in [K]$, noting that  $\alpha=D/K$ does not depend on $\{\beta_i\}_{i\in [K]}$, and ${\sum_{i\in [K]}\beta_i = M}$. 
This readily implies that $\sum_{i\in [K]} (\beta_i/(\alpha+\beta_i))\leq (M/(D+M))K$.


\section{Proof of Theorem~\ref{thm:LIPIRAch}}\label{sec:Ach}
In this section, we present a linear IPIR protocol, referred to as \emph{Group-and-Code}, for all ${K,D,M}$ such that ${(D/R+M/R)\mid K}$, where ${R:= \gcd(D,M)}$, and show that this protocol achieves the rate $(D+M)/K$. 
The Group-and-Code protocol consists of three steps described below. 

For ease of notation, we define ${d:=D/R}$,  ${m:=M/R}$, ${T:=d+m}$, and ${P:=K/T}$. 

\vspace{0.125cm}
\textbf{Step~1:} 
First, the user randomly partitions the message indices $1,\dots,K$ into ${P}$ groups $\mathcal{I}_1,\dots,\mathcal{I}_P$, each of size ${T}$ as follows: 
(i) $R$ (out of $P$) groups are chosen at random, and each of these groups is filled with ${d}$ randomly chosen message indices from $\mathrm{W}$ and $m$ randomly chosen message indices from $\mathrm{S}$; and 
(ii) the remaining ${P-R}$ groups are randomly filled with the remaining ${K-D-M}$ message indices from ${[K]\setminus (\mathrm{W}\cup\mathrm{S})}$. 
Next, the user constructs $d$ arbitrary length-$T$ row-vectors $\mathrm{v}_1,\dots,\mathrm{v}_{d}$ with entries from $\mathbbmss{F}_q$ such that the matrix ${\mathrm{V}:=[\mathrm{v}_{1}^{\mathsf{T}},\dots,\mathrm{v}_{d}^{\mathsf{T}}]^{\mathsf{T}}}$ generates a ${[T,d]}$ MDS code over $\mathbbmss{F}_q$. 
The user then sends ${\mathcal{I}_1,\dots,\mathcal{I}_P}$ and ${\mathrm{v}_1,\dots,\mathrm{v}_{d}}$ as the query $\mathrm{Q}^{[\mathrm{W},\mathrm{S}]}$ to the server. 

\vspace{0.125cm}
\textbf{Step~2:} Given the query $\mathrm{Q}^{[\mathrm{W},\mathrm{S}]}$, for each ${k\in [P]}$ and each ${l\in [d]}$, the server computes ${\mathrm{Z}_{k,l} = \sum_{j\in [T]} v_{l,j} \mathrm{X}_{i_{k,j}}}$, where $\mathcal{I}_{k}=\{i_{k,1},\dots,i_{k,T}\}$ and $\mathrm{v}_l = [v_{l,1},\dots,v_{l,T}]$, and sends $\{\mathrm{Z}_{k,l}\}_{k\in [P],l\in [d]}$ back to the user as the answer $\mathrm{A}^{[\mathrm{W},\mathrm{S}]}$.  

\vspace{0.125cm}
\textbf{Step~3:} Given the answer $\mathrm{A}^{[\mathrm{W},\mathrm{S}]}$ and the side information $\mathrm{X}_{\mathrm{S}}$, the user recovers their demand $\mathrm{X}_{\mathrm{W}}$ as follows. 
Without loss of generality, assume that 
${\mathcal{I}_1,\dots,\mathcal{I}_R}$ contain messages indices from ${\mathrm{W}\cup\mathrm{S}}$.
For each ${k\in [R]}$, let ${\mathrm{W}_{k}\subset \mathcal{I}_k}$ and ${\mathrm{S}_{k}\subset \mathcal{I}_k}$ be such that 
(i) ${\mathrm{W}_{k}\subseteq \mathrm{W}}$ and ${\mathrm{S}_{k}\subseteq \mathrm{S}}$, and 
(ii) ${\mathcal{I}_k = \mathrm{W}_k\cup \mathrm{S}_k}$.
For each ${k\in [R]}$ and each ${l\in [d]}$, the user computes ${\tilde{\mathrm{Z}}_{k,l}}$ by subtracting off the contribution of the $m$ side information messages $\mathrm{X}_{\mathrm{S}_k}$ from ${\mathrm{Z}_{k,l}}$, i.e., 
${\tilde{\mathrm{Z}}_{k,l} = \sum_{j\in \mathcal{J}_k} v_{l,j}\mathrm{X}_{i_{k,j}}}$, where ${\mathcal{J}_k\subset [T]}$ is such that ${\mathrm{W}_{k} = \{i_{k,j}: j\in \mathcal{J}_k\}}$.
For each ${k\in [R]}$, the user then recovers the $d$ demand messages ${\mathrm{X}_{\mathrm{W}_k}}$ from ${\tilde{\mathrm{Z}}_{k,1},\dots,\tilde{\mathrm{Z}}_{k,d}}$ by solving a system of linear equations.

The rate of the Group-and-Code protocol is equal to ${(D+M)/K}$.
Note that for any realization $(\mathrm{W},\mathrm{S})$, the answer $\mathrm{A}^{[\mathrm{W},\mathrm{S}]}$ consists of $L:=Pd $ ($=Kd/(d+m)=KD/(D+M)$) linear combinations $\{\mathrm{Z}_{k,l}\}_{k\in [P],l\in [d]}$.
It is also easy to see that $\{\mathrm{Z}_{k,l}\}_{k\in [P],l\in [d]}$ are linearly independent combinations of the messages. 
This implies that all realizations of $\mathbf{A}^{[\mathbf{W},\mathbf{S}]}$ consist of $L$ linearly independent combinations of the messages. 
Thus, $H(\mathbf{A}^{\mathbf{W},\mathbf{S}}) = LB$ (as discussed in Section~\ref{sec:SN}), where $B=H(\mathbf{X}_i)$ for all $i\in [K]$.
Since $H(\mathbf{X}_{\mathbf{W}}) = DB$, then the rate of this protocol is equal to ${DB/(LB) = D/L = (D+M)/K}$.  

It is also easy to see that the recoverability condition is satisfied.  
Fix an arbitrary $k\in [R]$. 
Note that ${\tilde{\mathrm{Z}}_{k,1},\dots,\tilde{\mathrm{Z}}_{k,d}}$ are linear combinations of the demand messages $\mathrm{X}_{\mathrm{W}_k}$.
Note also that the coefficient vectors of these linear combinations are the rows of a ${d\times d}$ submatrix of $\mathrm{V}$, and every ${d\times d}$ submatrix of $\mathrm{V}$ is full-rank because $\mathrm{V}$ generates a ${[T,d]}$ MDS code. 
Thus, ${\tilde{\mathrm{Z}}_{k,1},\dots,\tilde{\mathrm{Z}}_{k,d}}$ are $d$ linearly independent combinations of the $d$ demand messages $\mathrm{X}_{\mathrm{W}_k}$.

To prove that the individual privacy condition is satisfied, we need to show that ${\mathbb{P}(i\in \mathbf{W}|\mathbf{Q}^{[\mathbf{W},\mathbf{S}]}=\mathrm{Q}^{[\mathrm{W},\mathrm{S}]}) = D/K}$ for all ${i\in [K]}$. 
Fix an arbitrary ${i\in [K]}$.
Let ${k\in [P]}$ be such that ${i\in \mathcal{I}_{k}}$. 
From the description of the protocol, it is easy to see that ${\mathbb{P}(i\in \mathbf{W}|\mathbf{Q}^{[\mathbf{W},\mathbf{S}]}=\mathrm{Q}^{[\mathrm{W},\mathrm{S}]})}$ is equal to the probability that the $k$th group is one of the $R$ groups that are randomly chosen at first and the message index $i$ is one of the $d$ demand message indices that are placed in the $k$th group (i.e., the message index $i$ belongs to $\mathrm{W}_{k}$). 
The probability that the $k$th group is one of the $R$ chosen groups is ${R/P = (D+M)/K}$, and given that the $k$th group is one of the $R$ chosen groups, the probability that the message index $i$ belongs to $\mathrm{W}_{k}$ is ${d/T = D/(D+M)}$. 
Thus, ${\mathbb{P}(i\in \mathbf{W}|\mathbf{Q}^{[\mathbf{W},\mathbf{S}]}=\mathrm{Q}^{[\mathrm{W},\mathrm{S}]}) = (R/P)\times (d/T) = D/K}$.

\bibliographystyle{IEEEtran}
\bibliography{PIR_PC_Refs}

\end{document}